\documentclass[fleqn,10pt]{wlscirep}
\usepackage[utf8]{inputenc}
\usepackage[T1]{fontenc}
\usepackage{subcaption}
\usepackage{hyperref}
\usepackage{comment}
\usepackage{graphicx}
\usepackage{amsmath}
\usepackage[absolute,overlay]{textpos}

\usepackage{amsmath}
\title{Quantifying community evolution in temporal networks}

\author[1,*]{Peijie Zhong}
\author[2]{Cheick Ba}
\author[1]{Ra\'{u}l Mondrag\'{o}n}
\author[1]{Richard Clegg}
\affil[1]{School of Electronic Engineering and Computer Science, Queen Mary University of London, London E1 4NS, United Kingdom}
\affil[2]{Pometry, London EC2A 2BB, United Kingdom}
\affil[*]{p.zhong@qmul.ac.uk}


\begin{abstract}
When we detect communities in temporal networks it is important to ask questions about how they change over time. Adjusted mutual information (AMI) has been used to measure the similarity of communities when the nodes on a network do not change. We propose two extensions, namely, Union-Adjusted Mutual Information (UAMI) and Intersection-Adjusted Mutual Information (IAMI). UAMI and IAMI evaluate the similarities of community structures when nodes are added or removed. Experiments show that these methods are effective in dealing with temporal networks with the changes in the set of nodes, and can capture the dynamic evolution of community structure in both synthetic and real temporal networks. This study not only provides a new similarity measurement method for network analysis but also deepens the understanding of community change in complex temporal networks.
\end{abstract}
\begin{document}

\flushbottom
\maketitle
%
%
\thispagestyle{empty}
\begin{textblock*}{10cm}(0.5cm,0.5cm) 
{\small * This paper has been accepted to publish on Scientific Reports.}
\end{textblock*}
\section{Introduction}
An important problem in network science is understanding how communities form in networks. Communities within these networks do not remain static: they expand, merge, or dissolve, reflecting the underlying changes in relationships and interactions~\cite{tardelli2024temporal,greene2010Tracking,chen2010detecting,du2015tracking}. We might hypothesise that some networks have relatively stable and long-lasting relationships but other networks might form only transient communities, but how can this be measured?
A community here is a partition of a network into (usually) distinct sets of nodes with connections more common within communities. A common measure used on such a partition is modularity~\cite{newman2003structure, newman2006modularity, fortunato2007resolution}, though it has attracted some criticism~\cite{peixoto2023descriptive}. When we consider a network that changes in time we can choose to analyse the communities only within a time window and ask how much communities change between such windows. This requires us to rigorously compare two different partitions of a network and to give a measure of how similar those partitions are. This measure must allow for the fact that nodes can leave or join the network. 

A well-known method for comparing two such partitions is adjusted mutual information (AMI) but this measure assumes that both networks have the same set of nodes. The primary challenge addressed in this research is the accurate comparison of community structures in networks where the set of nodes varies over time. This study proposes variants of AMI: UAMI and IAMI which compares two networks with different node sets either by considering the union or intersection of the node sets. We show by using artificial models that these measure different aspects of how communities change over time and demonstrate their utility on artificial networks with known characteristics. Following this we use real temporal network data sets which we split into different time windows. We demonstrate how graph statistics can be used to pick an appropriate time window for the task. The methods accurately recover the known properties of artificial data.  On real-world datasets, however, community stability is generally brief—shorter than in our synthetic models—and varies across different cases.
\section{Related work}
The focus of this paper is on how communities change rather than specifically on how to partition a network into communities. Here we begin with a basic assumption, commonly made, that a community detection algorithm partitions all nodes in a network into a set of non-overlapping communities. Community detection is far from a solved problem. The problem can be divided into inferential methods and descriptive methods~\cite{peixoto2023descriptive}. Inferential methods attempt to fit parameters in some predetermined model that generates communities. Descriptive methods attempt to partition the network to maximise some measure of how good that partition is at creating communities. Modularity~\cite{newman2006modularity} is commonly used for the latter as it attempts to measure the proportion of links that start and end in the same community rather than that which would be expected by a null model. However, modularity suffers from a resolution limit~\cite{fortunato2007resolution}, causing small but meaningful communities to be merged in large networks. To mitigate this problem, previous research introduced modularity density to better capture within-community link compactness, and adopted z-score to evaluates the statistical significance of modularity scores against a configuration-model ensemble~\cite{chen2014community, atsushi2016zscore}. In this paper, the measures we use do not rely on whether descriptive or inferential techniques are available. We use the Louvain algorithm to detect communities for our investigation into real-world data because it is well-known, relatively quick to run, and simple to understand. However, the core contribution of the UAMI and IAMI and their use to measure how fast communities change in time would not change if a different community detection method were used~\cite{raghavan2007near,rosvall2008maps}. For the work on artificial networks, we simply assume that the communities exist (community membership is generated directly from the model) and that some algorithm could in principle detect them. 

The dynamic characteristics of dynamic communities play an important role in the study of the evolution of complex systems, which provides a bridge for the interaction between micro and macro structures in complex systems. To effectively capture these evolving structures over time, researchers commonly use the snapshot model to analyse dynamic networks. Each snapshot corresponds to a static network, and the community structure on each snapshot is obtained by using the static community detection algorithm, then the evolution of the community structure over time can be analysed by comparing the structure on different snapshot. Thus, the choice of the size of time window is important. A small time window can capture short-term fluctuations~\cite{krings2012effects}, but it is subject to noise interference. A large time window can smooth out short-term fluctuations, but it will lose local details~\cite{cleveland2013smoothing}. The paper~\cite{peel2015detecting} applies the changing point detection method to locate the moments when the network structure undergoes significant changes, providing a data-driven basis for the division of the time window. The paper~\cite{cazabet2021data} utilises the principle of data compression and selects the dynamic network expression form that achieves the best compromise between representation efficiency and interpretability. Paper~\cite{peixoto2017modelling} proposed a model based on dynamic stochastic block model that can non-parametically infer the number of communities and their evolution time scale in a continuous-time or non-stationary background, without the need to set the window length in advance. The paper~\cite{delvenne2010stability} discusses the stability of the community structure in networks and proposes a stability measurement method based on the Markov process. This method reveals the community structure characteristics of the network at different resolutions through time-scale analysis. 

Based on existing studies, we improved existing techniques and compared the similarity of community structure of temporal networks at different time points on a one-to-one basis. A direct approach to comparing the similarity of networks at the mesoscale level is comparing the similarity of the partition of nodes. Normalised Mutual Information (NMI)~\cite{danon2005comparing} is a widely used method to measure the similarity of partitions, by comparing the similarity of two partitions by measuring information gain. Recent studies have pointed out that the NMI is biased, and its variants~\cite{vinh2009information, amelio2015normalized} and asymmetric normalise technique~\cite{jerdee2023normalized} are proposed to solve the problem. In the context of overlapping communities, where nodes can have more than one community labels, Overlapping NMI can be used~\cite{mcdaid2011normalized, lancichinetti2009detecting} to measure the similarity in community structures. In addition, partition edit distance (PED)~\cite{aynaud2010static} is an extension of graph edit distance that defines operands to manipulate the partition of nodes. However, it relies on one-to-one matching between communities, and the high computational overhead also limits the application of this method on large-scale networks. 
\section{Methods}
Comparing network partitions is a fundamental task in network science and community detection. Mutual Information (MI) and its extensions are widely used metrics due to their ability to effectively quantify the similarity between partitions. However, mutual information assumes that the partitions being compared share an identical set of nodes, which often is not the case in real-world scenarios where networks may have differing node sets due to growth, decay, or entirely different origins. This limitation becomes evident when attempting to compare partitions of networks with varying node compositions, rendering traditional mutual information inadequate for such analyses. Recognising this gap, we propose an approach that explicitly compares network partitions with different node sets by introducing two novel metrics: Union-AMI (UAMI) and Intersection-AMI (IAMI). This approach is motivated by the practical need to assess the similarity between partitions in networks where node addition or deletion is common, such as in temporal networks or comparative studies across different systems.

In our framework, UAMI considers the union of the node sets from the partitions being compared, effectively incorporating all nodes present in either partition. This allows for a comprehensive evaluation of similarity that accounts for the entirety of both networks. Conversely, IAMI focuses on the intersection of the node sets, assessing similarity based solely on the nodes common to both partitions.
\subsection{Temporal networks model}
We adopt the snapshot model~\cite{holme2019temporal} of a sequential network to study the dynamic evolution of networks in time dimension. The model works by dividing the entire temporal network into a series of discrete time segments, each corresponding to a snapshot of the temporal network. This method allows us to observe changes in the structure of the network at different points in time and analyse the dynamic behaviour of nodes and edges. Here we consider our graph to be constructed from a series of events (edges) that connect two nodes and have a given time of occurrence. A temporal network that is observed between $t_0$ and $t_{\max}$ can be represented by a series of snapshots $(G_1, G_2, \dots, G_n)$ where each snapshot contains only the edges in the time window $T_i\in [t, t+\tau)$, denoted by , where $\tau$ is the size of the time window. Each snapshot $G_i=(V_i,E_i)$ represents the subgraph in time window $T_i$. $V_i$ are the nodes $E_i$ are the edges are active in time window $T_i$. We can also modify this definition slightly to allow snapshots where the time windows overlap. 

Choosing the right window size is critical to building a snapshot model of a dynamic network. A small time window will result in sparse graphs within snapshots as the number of edges within the window will be small. A large time window smooths out short-term fluctuations, thereby losing local information. Although several methods for extracting timescales in dynamic data sets exist already~\cite{darst2016detection,peel2015detecting,cazabet2021data,peixoto2017modelling}, given that our problem is choosing a reasonable time window to build a network with meaningful community structure, we analyse the modularity of the network under different windows and the proportion of the largest connected component to select the appropriate time window. Intuitively, in a snapshot model constructed with a reasonable time window, the modularity of each time slice should be large, indicating that there is a meaningful community structure in the network at this time. The research~\cite{fortunato2007resolution} indicates that modularity is not fair when comparing networks of different sizes and densities, and this is known as the resolution limit of modularity. Therefore, we eliminate the bias in modularity caused by networks of different sizes and densities by comparing the z-scores~\cite{chen2015new,chen2018network,atsushi2016zscore} obtained from the null model~\cite{newman2003structure}. Given a network, $G = (V, E)$ where $V$ is the set of nodes, $E$ is the set of edges and the degree of node $i$ is $d_i$. The configuration model randomly reconnects edges in the network by keeping the degree distribution of nodes constant to generate a network with the same degree sequence but randomly connected $G' = (V, E')$. For the original network $G$ and each generated random network $G'_m (m = 1, 2, \dots, M )$, the same community detection algorithm is used to partition the community, and the corresponding modularity $Q$ is calculated. For the $M$ random networks $G'_1, G'_2, \dots, G'_M$, calculating the modularity of the corresponding $Q'_1, Q'_2, \dots, Q'_M$. Through these modularity values, we can get the distribution characteristics of the modularity of random networks, including the expected value $\mu_{Q'}$ and the standard deviation $\sigma_{Q'}$, where
\begin{equation*}
    \mu_{Q^{\prime}}=\frac{1}{M} \sum_{m=1}^M Q^{\prime}_m;\quad \sigma_{Q^{\prime}}=\sqrt{\frac{1}{M-1} \sum_{m=1}^M\left(Q^{\prime}_m-\mu Q^{\prime}\right)^2}.
\end{equation*}
Thus, we can compare the observed modularity of the network with the modularity of this group of random networks to determine whether the network does have a strong community structure, or whether the high modularity of the network is caused by the low density of edges. We measure the significance of modularity using a z-score. It is defined by: $Z=\frac{Q_{\mathrm{obs}}-\mu_{Q^{\prime}}}{\sigma_{Q^{\prime}}}$, where $Q_{\mathrm{obs}}$ is the modularity of the original network.

\subsection{Mutual information and the extensions}
\begin{figure*}
\centering
\includegraphics[width=\linewidth]{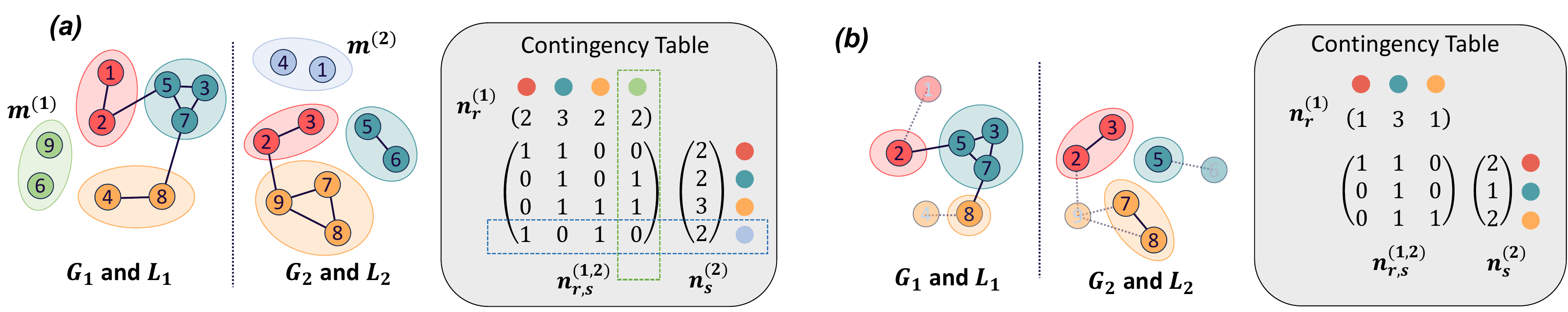}
\caption{The diagram shows the principles of union extension and intersection extension respectively. In the two networks, the nodes labelled 6 and 9 appear in $G_1$ but are removed in $G_2$. The nodes labelled 1 and 4 are new in $G_2$. (a) The union extension compares the union of two node sets and adds a community labelled with $m^{(1)}$ in the partition of $G_1$, including nodes 6, and 9. A new community labelled $m^{(2)}$ is added to the partition of $G_2$, including nodes 1 and 4. (B) The intersection extension compares the intersection of two node sets, ignoring the added/removed nodes. In the example, nodes 1, 4, 6 and 9 are excluded when comparing the similarity.} 
\label{fig:demo}
\end{figure*}
Recalling the definition of the mutual information in a static graph context, given a network $G=(V,E)$ with community structures, let $N$ be the number of nodes in $G$. A community detection algorithm gives each node in $V$ a label. Let $L_1=\{1,2,\dots,r\}$ and $L_2=\{1,2,\dots,s\}$ be two such labellings where the actual value of the label is arbitrary (so two labellings are considered the same if they partition $V$ into the same groups even if the actual labels are different). Let $n_r^{(1)}$ be the number of nodes labelled $r$ by $L_1$ and $n_s^{(2)}$ be the number of nodes labelled $s$ in the second labelling $L_2$ and $n^{(1,2)}_{r,s}$ be the number of nodes labelled $r$ in $L_1$ and labelled $s$ in $L_2$. Then we have the probability $P(L_1,r)$ represents the proportion of nodes labelled $r$ in $L_1$, $P(L_2,s)$ represents the proportion of nodes labelled $s$ in $L_2$ and $P(L_1,r,L_2,s)$ represents the proportion of nodes labelled $r$ in $L_1$ and labelled $s$ in $L_2$:
\begin{equation*}
P(L_1,r)=\frac{n^{(1)}_r}{N}; P(L_2,s)=\frac{n^{(2)}_s}{N};P(L_1,r,L_2,s)=\frac{n^{(1,2)}_{r,s}}{N}.
\end{equation*}
The entropy of the labelling scheme $L_i$ where $i \in \{1,2\}$ is defined by 
\begin{equation*}
H(L_i)=\sum_r -P(L_i,r)\log P(L_i,r).
\end{equation*}
The mutual information between the two labelling schemes is
\begin{equation*}
I(L_1;L_2)=\sum_s \sum_r P(L_1,L_2)\log \left(\frac{P(L_1,L_2)}{P(L_1)P(L_2)}\right)=\sum_r \sum_s \frac{n^{(1,2)}_{r,s}}{N}\log\left(\frac{Nn^{(1,2)}_{r,s}}{n^{(1)}_r n^{(2)}_s}\right).
\end{equation*}
One extension of mutual information is the Normalised Mutual Information (NMI), which is obtained by normalising the mutual information by the mean of the entropies of the two labelings. Depending on the application, either the arithmetic mean or the geometric mean may be used for normalisation. In this work, we adopt the arithmetic mean. Recent studies have shown that NMI can be biased by the number of communities, which means that NMI tends to give high scores to communities with a large number of small partitions~\cite{vinh2009information, jerdee2023normalized, mccarthy2019metrics}. Thus, adjusted mutual information (AMI)~\cite{vinh2009information} is proposed to correct for this bias, pointed out that the bias can be eliminated by subtracting the expected value under the null model configuration, the expectation term can be expressed by:
\begin{equation}
    \mathbb{E}[\operatorname{MI}]=\sum_{r=L_1} \sum_{s=L_2} \sum_{n=n_{r,s}^{\min}}^{n_{r,s}^{\max}} \frac{\binom{n_r^{(1)}}{n} \binom{N-n_r^{(1)}} {n_s^{(2)}-n}} {\binom{N}{n_s^{(2)}}}\times \frac{n}{N}\log\left(\frac{nN}{n_r^{(1)}n_{s}^{(2)}}\right).
\end{equation}
Then the expression of AMI is $\frac{2(\operatorname{MI}-\mathbb{E}[\operatorname{MI}])}{H(L_1)+H(L_2)-\mathbb{E}[\operatorname{MI}]}$. The AMI takes a value of 1 when the two partitions are identical and 0 when the MI between two partitions equals the value expected due to chance alone. In the following expansion, we will apply the same operation to eliminate the bias caused by the number of communities.
\subsubsection{Union-AMI}
Our first proposal is to align two node sets by taking the union set of two partitions. Formally, given two networks $G_1=(V_1,E_1)$ with labelling scheme $L_1$, $G_2=(V_2,E_2)$ with labelling scheme $L_2$. $n^{(1)}_r$ is the number of nodes labelled $r$ in $L_1$ and $n^{(2)}_s$ is the number of nodes labelled $s$ in $L_2$ and $n^{(1, 2)}_{r,s}$ is the number of nodes labelled $r$ in $L_1$ and labelled $s$ in $L_2$. Nodes may be added or removed from the network between $G_1$ and $G_2$ thus $V_1$ and $V_2$ are two sets that are usually not equal and each may contain nodes not in the other, thus We have $V_1-V_2$ represents nodes in $V_1$ but not in $V_2$ and $V_2-V_1$ is nodes that are in $V_2$ but not in $V_1$. Let $N_U=|V_1\cup V_2|$ be the total number of nodes. The union extension solves the problem of the inconsistent number of nodes in the two networks by pretending that nodes in $G_1$ but not in $G_2$ or vice versa are actually present and in some ``virtual" community. We pretend these nodes are present in a community labelled $m^{(1)}$ in the labelling scheme $L_1$ and a community labelled $m^{(2)}$ in the labelling scheme $L_2$ (Fig~\ref{fig:demo}(a)). The mutual information between the two labelling schemes is 
\begin{equation}
\begin{aligned}
     I_{U}(L_1;L_2)=&\sum_{r\neq m^{(1)}}\sum_{s \neq m^{(2)}} \frac{n^{(1,2)}_{r,s}}{N_U}\log\left(\frac{N_U n_{r,s}^{(1,2)}}{n^{(1)}_r n^{(2)}_s}\right)\\
    &+\sum_{s}\frac{n^{(1,2)}_{m^{(1)},s}}{N_U}\log\left(\frac{N_U n^{(1,2)}_{m^{(1)},s}}{|V_2-V_1|n^{(1)}_r}\right)\\
    &+\sum_{r}\frac{n^{(1,2)}_{r,m^{(2)}}}{N_U}\log\left(\frac{N_U n^{(1,2)}_{r,m^{(2)}}}{|V_1-V_2|n^{(2)}_s}\right). 
\end{aligned}
\label{eq:mutual-UNMI}
\end{equation}
The first term represents the mutual information of the partition of common nodes. The second term represents the mutual information calculated by the virtual community added in $L_1$ and all communities in $L_2$, which measures the complexity of node flow during the transition from $G_1$ to $G_2$, and a smaller value represents a more complex flow of nodes in the virtual community, indicating that they have been assigned to multiple different communities during the evolution process. The third term represents the new virtual community in $L_2$ and the mutual information of all communities in $L_1$, and it measures the complexity of node computations during the transition from $G_1$ to $G_2$. The mutual information is normalised by the arithmetic average of entropy in two labelling schemes, which are for
$i=\{1,2\}$ and $j=2$ if $i=1$ or $j=1$ if $i=2$,
\begin{equation}
H_U(L_i)=-\sum_{r\neq m^{(i)}} \frac{n^{(i)}_{r}}{N_U}\log\left(\frac{n^{(i)}_r}{N_U}\right)-\frac{|V_i-V_j|}{N_U}\log\left(\frac{|V_i-V_j|}{N_U}\right),
\label{eq:entropy-UNMI}
\end{equation}
where $|\cdot|$ refers to the cardinality of a set. The first term represents the entropy of the original community label after introducing the virtual community, and the second term represents the entropy introduced due to the virtual community. Subsequently, by subtracting the average value of the mutual information that can be generated by random partitioning under the null model assumption. We obtained UAMI. The expression of the expectation term is as follows:
\begin{equation}
\mathbb{E}_U[\operatorname{MI}]=\sum_{r=L_1\cup \{m^{(1)}\}} \sum_{s=L_2\cup \{m^{(2)}\}} \sum_{n=n_{r,s}^{\min}}^{n_{r,s}^{\max}} \frac{\binom{n_r^{(1)}}{n} \binom{N_U-n_r^{(1)}}{n_s^{(2)}-n}}{\binom{N_U}{n_s^{(2)}}}\times \frac{n}{N_U}\log\left(\frac{nN_U}{n_r^{(1)}n_{s}^{(2)}}\right).
\end{equation}
Combining Eq.~\ref{eq:mutual-UNMI} and Eq.~\ref{eq:entropy-UNMI}, then the expression of UAMI is $\frac{2(I_I(L_1;L_2)-\mathbb{E}_U[\operatorname{MI}])}{H_I(L_1)+H_I(L_2)-\mathbb{E}_U[\operatorname{MI}]}$.
\subsubsection{Intersection-AMI}
The second proposal is to align the node set by taking the intersection set of the partitions, which evaluates the overlap in information between two community structures. The intersection extension can be interpreted as a measure of coherence or consistency between the two structures. A higher value in the intersection extension means that the shared community structure captures a significant portion of the information from both original structures. It emphasises the regions where the structures agree, making it useful for identifying core similarities. Here, the number of common nodes in $G_1$ and $G_2$ is denoted by $N_I=|V_1\cap V_2|$ and let $q_r^{(1)}$ be the number of nodes that are labelled $r$ in $L_1$ and are also in $G_2$, and $q_s^{(2)}$ be the number of nodes that are labelled $r$ in $L_2$ and are also in $G_1$ and $q_{r,s}^{(1,2)}$ is the number of nodes labelled $r$ in $L_1$ and $s$ in $L_2$ and they appears in both $G_1$ and $G_2$ (Fig.~\ref{fig:demo}(b)). The mutual information is
\begin{equation}
I_I(L_1;L_2)=\sum_{r,s}\frac{q^{(1,2)}_{r,s}}{N_I}\log\left(\frac{q^{(1,2)}_{r,s}N_I}{q_r^{(1)}q_s^{(2)}}\right).     
\label{eq:mutual-INMI}
\end{equation}
The entropies of the two partitions of the networks are for $i \in \{1,2\}$:
\begin{equation}
    H_I(L_i)=-\sum_r \frac{q_r^{(i)}}{N_I}\log\left(\frac{q_r^{(i)}}{N_I}\right).
\label{eq:entropy-INMI}
\end{equation}
Again, we subtract the expectation to eliminate the bias caused by the number of communities. When taking the intersection, the expectation term is expressed as:
\begin{equation}
    \mathbb{E}_I[\mathrm{MI}]=\sum_r \sum_s \sum_{q=q_{r,s}^{\min }}^{q_{r,s}^{\max }} \frac{\binom{q_r^{(1)}}{q}\binom{N_I-q_r^{(2)}}{q_s^{(2)}-q}}{\binom{N_I}{q_s^{(2)}}} \times \frac{q}{N_I} \log \left(\frac{qN_I }{q_r^{(1)} q_s^{(2)}}\right).
\end{equation}
Then the expression of IAMI is $\frac{2(I_I(L_1;L_2)-\mathbb{E}_I[\operatorname{MI}])}{H_I(L_1)+H_I(L_2)-\mathbb{E}_I[\operatorname{MI}]}$.

Because the intersection extension focuses only on nodes that occur in both $G_1$ and $G_2$ if the nodes in the intersection $V_1 \cap V_2$ remain in the same community then the intersection will be one irrespective of how many extra nodes are in $G_1$ and $G_2$. This invariance occurs because the IAMI is designed to measure the similarity based only on the set of nodes present in both partitions being compared, ignoring any nodes that are unique to one partition. However, the IAMI value changes when the nodes' community labels are shuffled. As the shuffle ratio increases, the community structure similarity, as captured by the IAMI, gradually decreases. This decrease is due to the reduction in mutual information and the corresponding increase in entropy, reflecting a loss of community structure coherence among shared nodes. 

\section{Results}

In this section, we apply our community structure similarity measure to synthetic temporal networks and multiple empirical data sets. In the synthetic temporal networks, we create a very simple synthetic model to generate communities which change quickly or slowly. We follow this with the investigation of five real-world networks. In addition, we replicated the methods of the relevant studies~\cite{wang2022quantifying}, demonstrating the reliability of our approach and enabling faster computations on the datasets (please refer to supplementary Fig. 9.
\subsection{Validation on synthetic data}
\begin{figure*}[h]
    \centering
    \includegraphics[width=0.95\linewidth]{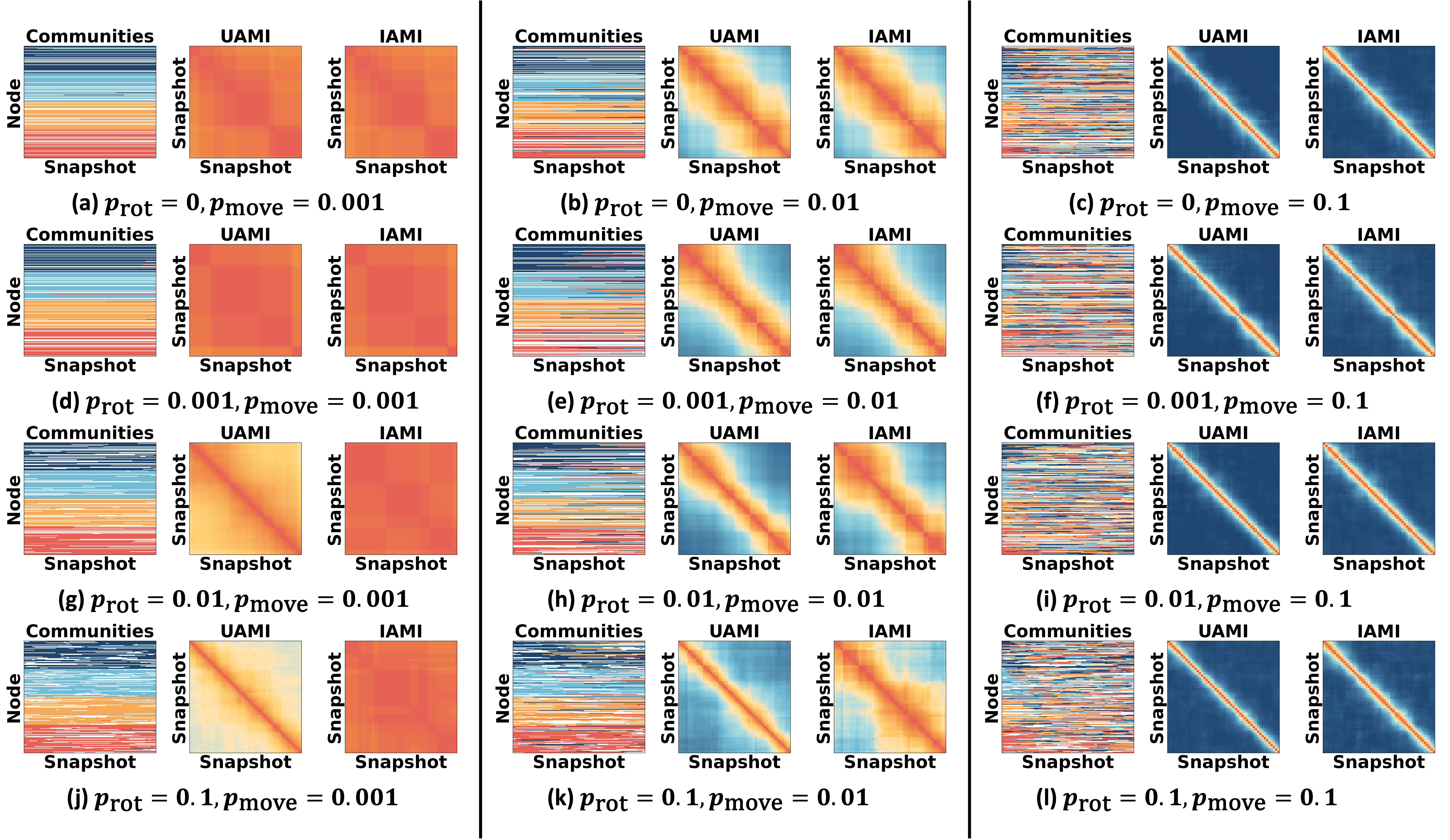}
    \caption{Example dynamic community structures with $n=400$ nodes and their pair-to-pair UAMI, IAMI measurement, where $p_{\mathrm{rot}} \in [0,1]$ is a parameter which, when high, means nodes leave/arrive in the network quickly (it increases from left to right), and $p_{\mathrm{move}} \in [0,1]$ is a parameter which, when high, means nodes move to new communities quickly (it increases from top to bottom). The first graph for each set of experiments represents the community assignments of nodes in different snapshots. We use four colours to represent the four communities specified in this experiment ( \includegraphics[scale=0.2]{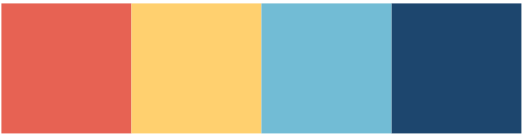} ). The two subsequent diagrams show the similarity measured by UAMI and IAMI in community structure between each pair of network slices. (0 \includegraphics[scale=0.2]{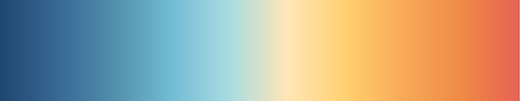} 1).}
    \label{fig:valid}
\end{figure*}
First, we create a candidate node pool with $N$ nodes, each of which is assigned an initial community label $C_i$ chosen at random from equiprobable communities (here we choose four). From these $N$ nodes, $n < N$ are chosen at random to form the initial network. Here we choose $N=500$ and $n=400$. A parameter $p_{\mathrm{rot}} \in [0,1]$ gives the probability a node will leave the network and be replaced by another node in the pool to keep the size of the network constant. A higher value of $p_{\mathrm{rot}}$ means nodes swap between the network and the pool often (they keep a memory of their community when not in the network). Following this, at each iteration, every one of the nodes now in the network will change their community labels with probability $p_{\mathrm{move}} \in [0,1]$ picking one of the remaining three labels with equal probability. A higher value of $p_{\mathrm{move}}$ means communities are fast-moving and nodes change between them quickly. We consider values of $p_{\mathrm{rot}} \in \{0,0.001, 0.01, 0.03, 0.1, 0.3\}$ and values of $p_{\mathrm{move}} \in \{0,0.001, 0.01, 0.03, 0.1, 0.3\}$ and run the simulation for 50 iterations measuring the UAMI and IAMI for each. In Fig.~\ref{fig:valid}, we present the experimental results of UAMI and IAMI under representative $p_{\mathrm{rot}}$ and $p_{\mathrm{move}}$. For more experimental results at different values, please refer to the supplementary Fig. 6-8.

The results are shown in Fig.~\ref{fig:valid} along with a plot showing the membership of each community (with white showing nodes in the pool, not the network). Moving down the diagram means increasing values of $p_{\mathrm{rot}}$ (nodes move in and out of the network more rapidly). Moving rightward on the diagram means nodes move between communities more rapidly. For example, at the top left (Fig.~\ref{fig:valid}(a)) nodes stay in the network and change community very slowly. As expected all values of IAMI and UAMI are high (red). Conversely, at the bottom right (Fig.~\ref{fig:valid}(l)) nodes move in and out of the network rapidly and change community quickly and the UAMI and IAMI are both low except for the diagonal where the same or similar time periods are compared. The UAMI and IAMI measures degree on how similar communities are when we have a situation where the node set changes rapidly but the community labels stay relatively constant (Fig.~\ref{fig:valid}(j,k)) -- here IAMI finds that communities are similar because the community labels themselves have not changed much but UAMI finds a smaller degree of similarity because the actual node set could be quite different. We argue that in this case, both methods provide complementary information, neither is superior both are needed to get the full picture. In situations where communities change relatively slowly  (Fig.~\ref{fig:valid}(e,h))  we can see that the central band is thicker indicating slowly moving communities.  
\begin{table}[ht]
\centering
    \begin{tabular}{|l|r|r|r|r|}
    \hline
    Dataset & \# Nodes & \# Edges & \# Events & Time-span \\
    \hline
    Email-EU-core & 986 & 24,929 & 332,334 & 10/2003 -- 4/2005 \\
    Math Overflow & 24,818 & 239,978 & 506,550 &  9/2009 -- 3/2016 \\ 
    arXiv Hep-Th & 16,959 & 1,194,440 & 2,322,259 & 1/1993 -- 3/2003 \\
    SubReddit & 53,018 & 207,636 & 510,787 & 1/2014 -- 4/2017 \\
    NFT trading & 532,945 & 2,954,521 & 6,071,027 & 11/2017 -- 4/2021\\
    \hline
    \end{tabular}
    \caption{Summary of datasets. An event is a connection between two nodes at a given time. One or more events between two nodes count as a single edge.}
\label{tab:dataset}
\end{table}
\subsection{Real-world networks}
We now test the effectiveness of the proposed Union-AMI (UAMI) and Intersection-AMI (IAMI) metrics on several real-world temporal networks (See Tab.~\ref{tab:dataset}). The networks are all temporal with events connecting nodes occurring at well-defined times. In this experiment, we pick a window size for each network using the methods described in the supplemental information.
Among them, email-EU-core~\cite{paranjape2017motifs} is a mail interactive network of European research institutions, Math Overflow~\cite{paranjape2017motifs} is an interactive network of online question and answer platform, arXiv Hep-TH~\cite{ryan2015the} is a paper citation network of high energy physics section on arXiv. SubReddit~\cite{kumar2018community} is an interactive network between subreddits (a SubReddit is a community of users) on the Reddit platform, and NFT~\cite{nadini2021mapping} (non-fungible token) is an online transaction network. In the preprocessing, we removed the self-loops in the networks and trimmed the data where the network was sparse. For details on data sets and data preprocessing, see the supporting information. We use the Raphtory software~\cite{steer2023raphtory} to create windows of the correct size for each graph and partition nodes into communities using the Louvain algorithm~\cite{blondel2008fast}. For the relevant statistical results, please refer to supplementary Fig. 1-5 in the supporting information. We slide the window by ten percent of the window size and the process. We then look at the UAMI and IAMI for these sliding windows producing a two-dimensional heat map as with the synthetic data. 
\begin{figure}[t]
    \centering
    \includegraphics[width=\linewidth]{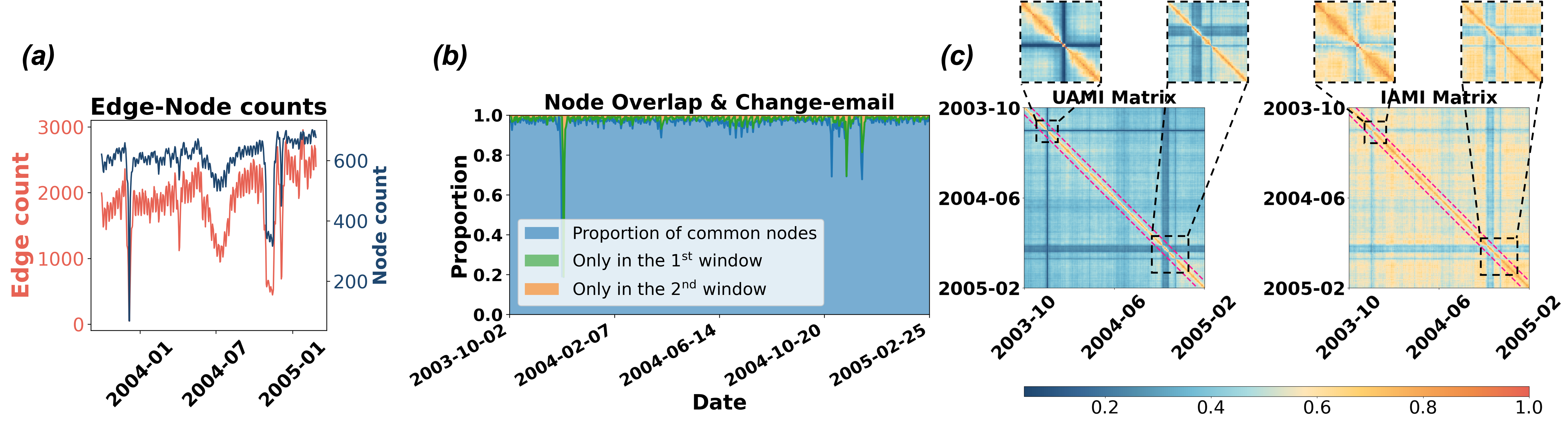}
    \caption{(a) shows edge and node counts. (b) shows the proportion of nodes appears only in the $1^{\text{st}}$ window, the $2^{\text{nd}}$ window, and appears in both windows. (c) shows the similarity in community measures in the email-EU-core network.}
    \label{fig:email-similarity}
\end{figure}
\begin{figure}[h]
\centering
\includegraphics[width=\linewidth]{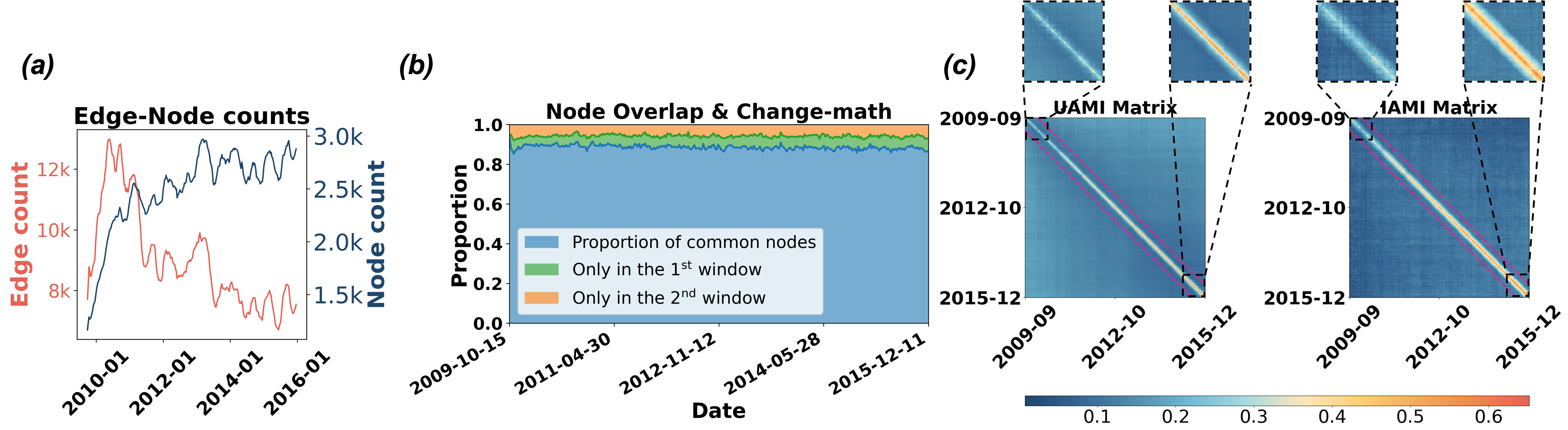}
    \caption{(a) shows edge and node counts. (b) shows the proportion of nodes appears only in the $1^{\text{st}}$ window, the $2^{\text{nd}}$ window, and appears in both windows. (c) shows the similarity in community measures in the Math Overflow network.}
    \label{fig:math-similarity}
\end{figure}
\begin{figure}[t]
\centering
\includegraphics[width=\linewidth]{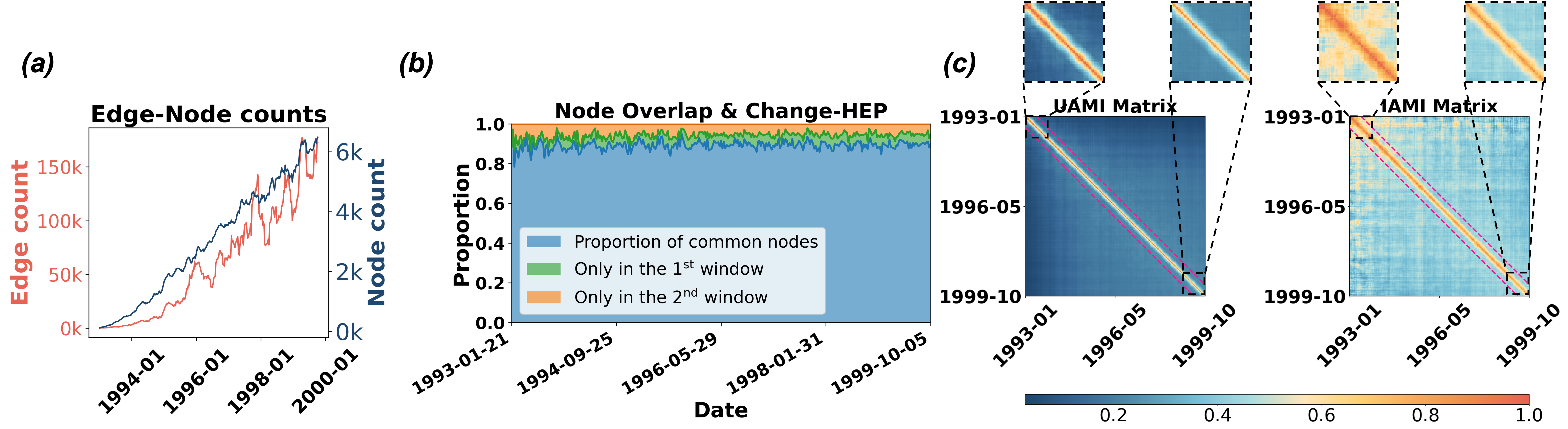}
    \caption{(a) shows edge and node counts. (b) shows the proportion of nodes appears only in the $1^{\text{st}}$ window, the $2^{\text{nd}}$ window, and appears in both windows. (c) shows the similarity in community measures in the arXiv HEP-TH network.}
    \label{fig:hep-similarity}
\end{figure}
\begin{figure}[h]
\centering
    \includegraphics[width=\linewidth]{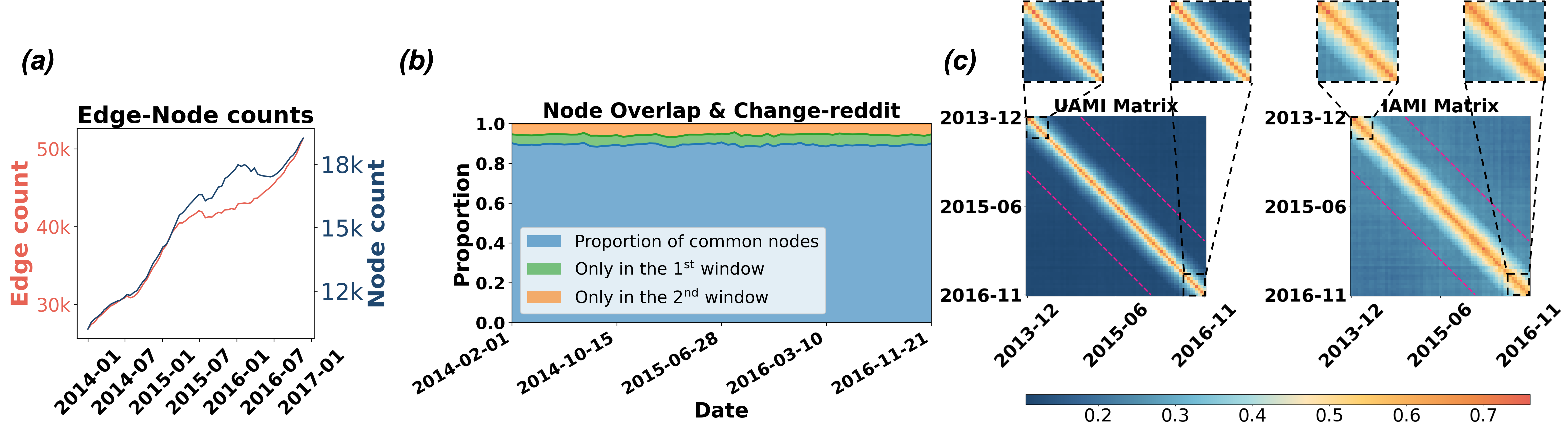}
    \caption{(a) shows edge and node counts. (b) shows the proportion of nodes appears only in the $1^{\text{st}}$ window, the $2^{\text{nd}}$ window, and appears in both windows. (c) shows the similarity in community measures in the subreddit network.}
    \label{fig:reddit-similarity}
\end{figure}
\begin{figure}[t]
\centering
    \includegraphics[width=\linewidth]{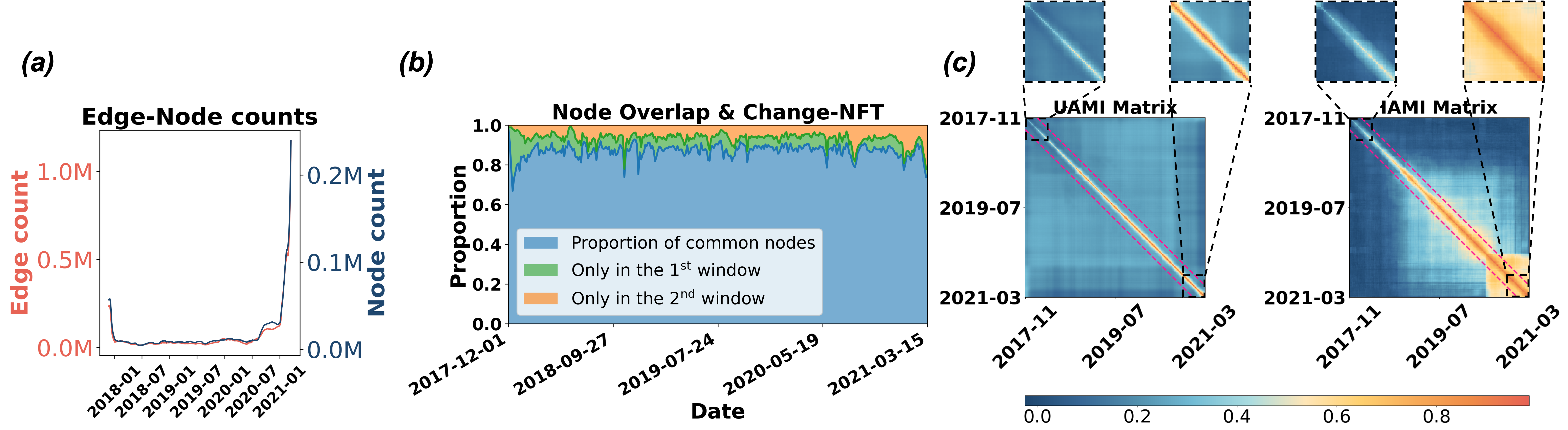}
    \caption{(a) shows edge and node counts. (b) shows the proportion of nodes appears only in the $1^{\text{st}}$ window, the $2^{\text{nd}}$ window, and appears in both windows. (c) shows the similarity in community measures in the NFT trade network.}
    \label{fig: nft-similarity}
\end{figure}

In the email-EU-core dataset (Fig.~\ref{fig:email-similarity}), which captures email communications within a large European research institution, we observed a significant decrease in UAMI readings around December 2003 and 2004, corresponding to the Christmas holiday periods. During the same time, IAMI readings showed a notable increase but the UAMI had a decrease (see zoom in). This suggests that while there were substantial changes in the overall network—reflected by the varying node sets due to personnel changes and reduced communication activity—the core group of staff who remained active maintained similar interaction patterns. This shows how the measures respond differently to a change in network membership and is consistent with the results on synthetic data. The IAMI is consistently higher because in this network the actual participants change greatly between windows as different members of the organisation become more or less active in sending and receiving emails. The network size remains constant over time and the thickness of the diagonal shows that communities are being formed at the same rate throughout.

For the Math Overflow dataset (Fig.~\ref{fig:math-similarity}), an online platform where users ask and answer mathematical questions, the results showed that communities are slowly forming but have not yet established stable structures. The open and fluid interaction model of the platform allows users to freely engage in discussions without strict group boundaries, making it challenging for strong communities to form. Although users change, both IAMI and UAMI show the main diagonal thickening over time indicating that as time continues users begin to form loosely connected communities. This leads to a slow but steady strengthening of community similarity among the core users. The graph here shows the whole history of the network so we might expect no coherent communities at the beginning of the data set. 

In the HEP-TH citation network (Fig.~\ref{fig:hep-similarity}), representing high-energy physics theory papers from the arXiv repository. The initial part of this data has been removed as the number of papers per year was very small and we can see the rate of nodes being added increases considerably. UAMI and IAMI both begin high and in particular at the beginning of the data set, where there are few papers, the IAMI is quite high. We can see the diagonal of the plot gets thinner for both as time goes on, perhaps indicating that as the number of contributors grows larger and larger that core forming a community is smaller by comparison. This reflects the dynamic nature of academic research, where the number of publications increases as time continues and evolving interests and expanding collaborations can fragment initially cohesive communities.

The dynamic community structure on the Reddit social platform (Fig.~\ref{fig:reddit-similarity}) is always relatively similar to several time slices adjacent to it in time, and this pattern does not change over time (It is reflected in the heatmap as a diagonal line with almost constant thickness), that is, there is neither the formation of communities nor the disappearance of communities. This section of the Reddit data comes from a midpoint in the network's history where the network is already well-established and not showing particularly rapid growth. The diagonal of the network is thicker than most others studied indicating that there is a persistent social structure compared with our other networks.

In the NFT transaction network (Fig.~\ref{fig: nft-similarity}), which records buying and selling activities in the NFT market, we observed that during the last year of the dataset, IAMI readings were significantly high, while UAMI decreased. The decrease in UAMI suggests that a large number of new participants entered the market, increasing the node count and altering the overall network structure, thus reducing similarity when considering all nodes. However, the high IAMI readings indicate that core buyers continued to interact predominantly with trusted users, maintaining consistent trading relationships. This stability among persistent participants led to more similar community structures within the core network. The findings align with previous research~\cite{nadini2021mapping} suggesting that despite market expansion, the NFT ecosystem is driven by a small number of active participants who engage in frequent transactions within established communities.

\section{Discussion}
The introduction of Union-AMI (UAMI) and Intersection-AMI (IAMI) addresses a significant gap in network analysis, particularly in the comparison of network partitions with differing node sets. Our experiments on both synthetic and real-world temporal networks demonstrate that these metrics offer robust, flexible, and meaningful assessments of partition similarity where traditional metrics fall short. One of the key findings from our experiments is the effectiveness of UAMI and IAMI in handling dynamic changes inherent in temporal networks. Traditional extension of MI assumes identical node sets between partitions, which is often not the case in real-world scenarios where nodes can join or leave the network over time. By explicitly considering the union or intersection of node sets, UAMI and IAMI accommodate these changes without resorting to artificial adjustments that can bias results.

One important aspect of our study is the interpretability of UAMI and IAMI. By building upon the foundational concepts of mutual information and adjusting them to account for node-set differences, these metrics retain the intuitive appeal of mutual information while extending its applicability. This makes them accessible tools for researchers who are already familiar with traditional network analysis techniques. Both UAMI and IAMI are necessary and one should not be preferred over the other as they give different insights, particularly when data sets have high node churn. In some circumstances, the answers they give are broadly the same but when the node overlap between two different windows is small compared with the network size, the difference between the two measures can be large. 

A surprise in our study was that none of the data sets we looked at showed long-lasting community structures which persisted over time. The networks studied rarely showed evidence for strong similarity in community structure lasting for any long proportion of the data set. This is somewhat surprising, especially for a network like the email-EU-core network which is the same group of people interacting over time. While our sample of five studied networks is too small to draw very general conclusions, it brings into question studies which have considered the communities formed by observing the networks statically with all these time windows compressed into one static graph structure. Of course this conclusion depends on the community detection algorithms used and also on the size of the persistent communities. It is possible that smaller stable communities exist within these networks but this stability is not reflected by the partition of the whole network into communities.  

In conclusion, our proposed metrics, UAMI and IAMI, enhance the toolkit available for network partition comparison by directly addressing the challenges posed by differing node sets. They offer a balance of theoretical rigour, practical applicability, and computational efficiency. There are practical considerations and limitations to keep in mind (obviously it does require more computational power than static analysis), however, the benefits they provide beyond simply looking at a static network are considerable.

\bibliography{refs}
\section*{Author contributions statement}
Peijie Zhong, Ra\'{u}l Mondrag\'{o}n and Richard Clegg designed the methods involved in this article. Peijie Zhong performed the experiments and analysed the data. Cheick Ba, Ra\'{u}l Mondrag\'{o}n and Richard Clegg advised on the analysis of the experiment data. 
\section*{Data availability}
The email-EU-core, Math Overflow, arXiv HEP-TH citation network and subReddit hyperlink network datasets are available at \url{https://snap.stanford.edu/data/index.html} and the NFT trading data is available at \url{https://osf.io/wsnzr/?view\_only=319a53cf1bf542bbbe538aba37916537}.
\section*{Code availability}
The code is available at \url{https://github.com/Peijie-Zhong/community-similarity}. The software Raphtory we used can be found at \url{https://github.com/Pometry/Raphtory}, the version number is v0.15.1.
\section*{Acknowledgement}
Thanks to Pometry for giving help and support with their Raphtory software that was used for many results in this paper. 
\section*{Additional information}
The authors declare no competing interests.
\end{document}